\documentclass[11pt,a4paper]{article}
\pdfoutput=1
\usepackage{graphicx}
\usepackage{amsmath}
\usepackage{amssymb}
\usepackage{url}
\usepackage{cite}
\graphicspath{{./}}
\begin{document}
\title{Scalar--photon coupling in cosmology: Radiation evolution, recombination, and CMB signatures}
\author{Yousef Bisabr\\[0.4em]\small Department of Physics, Shahid Rajaee Teacher Training University,\\\small Lavizan, Tehran 16788, Iran\\\small\texttt{y-bisabr@sru.ac.ir}}

\maketitle
\begin{abstract}
Interactions between a cosmological scalar field and the photon bath can modify the pre-recombination radiation history and leave signatures in cosmic microwave background (CMB) anisotropies. Their interpretation requires consistent evolution of the background and perturbations, together with an explicit account of the assumptions underlying photon thermodynamics.

We investigate the cosmological response to an effective scalar--photon-fluid coupling, distinguishing illustrative CMB signatures from observational constraints.

We derive the coupled background and linear fluid-level scalar perturbation equations and leading-order adiabatic initial relations for a canonical scalar field with an exponential potential. A radiation-era scaling branch sets the initial conditions; the subsequent departure from standard photon dilution is evolved dynamically. We implement the effective equations in CLASS, propagate the photon temperature through recombination under a Planckian-bath assumption, and calculate CMB and linear matter spectra. We evaluate conditional Planck 2018 TTTEEE-lite, low-multipole temperature and polarization, and lensing likelihoods, fixing the six standard cosmological parameters and profiling the calibration parameter $A_{\rm Planck}$. The photon collision hierarchy is treated phenomenologically rather than derived from a microscopic production operator.

Illustrative interacting solutions modify the expansion and recombination histories and produce scale-dependent changes in the CMB and matter spectra. The sampled conditional likelihood is shallow at small interaction strength. Its best sampled interacting point, $(\epsilon_i,\sigma)=(4\times10^{-5},0.10)$, is slightly less favored than the matched interaction-off $\Lambda$CDM reference ($\Delta\chi^2\simeq+0.30$). The fixed-baseline profile does not provide marginalized parameter constraints or a general confidence limit.

The effective coupling produces identifiable cosmological signatures in the implemented thermal-bath framework, but the conditional Planck comparison provides no preference for the interaction. The interpretation is limited by the assumed preservation of a thermal photon spectrum, the phenomenological treatment of photon collisions, and the distinction between thermal and gravitational photon-density normalization. These issues require further validation before definitive observational constraints can be claimed.
\end{abstract}
\medskip

\section{Introduction}
The spatially flat six-parameter $\Lambda$CDM model provides an accurate
description of the cosmic microwave background (CMB), the large-scale
expansion history and the growth of structure over a wide range of scales
\cite{peeb,hu,planck}. The final Planck analysis, in particular, showed that
temperature, polarization and lensing measurements are mutually consistent
with this minimal framework and determine the acoustic scale with very high
precision \cite{planck}. At the same time, the difference between the value of
the Hubble constant inferred within $\Lambda$CDM from early-Universe data and
direct late-time determinations has motivated extensive investigation of
physics operating before recombination \cite{riess2022,kamion}.

A prominent class of such proposals introduces an additional scalar component
that becomes dynamically relevant for a limited period before recombination.
In early dark energy (EDE) models, this additional energy density can increase
the pre-recombination expansion rate and reduce the sound horizon, thereby
allowing a larger CMB-inferred value of $H_0$
\cite{poulin2019,smith2020}. The same mechanism, however, also modifies the
evolution of perturbations and the matter power spectrum so that its viability
cannot be assessed from the background expansion alone
\cite{hill2020}. More generally, cosmological scalar fields may exchange
energy or momentum with other sectors as has been extensively studied in interacting dark energy and coupled quintessence models
\cite{amen,val,clemson,salvatelli,costa}. These studies emphasize the need to
evolve the background and linear perturbations of all interacting components
self-consistently.

The radiation sector provides a distinct way of probing non-standard
pre-recombination dynamics. In the standard adiabatic picture, the photon
energy density scales as $\rho_\gamma\propto a^{-4}$ and the CMB temperature
obeys $T_{\rm CMB}(z)=T_0(1+z)$. Photon production, absorption, conversion
or energy exchange between radiation and another cosmological component can
modify these relations. Such possibilities have been considered in decaying
vacuum models \cite{lima_vac,jetzer2011,jetzer2012}, phenomenological tests of photon non-conservation and violations of
distance duality
\cite{avgou,bassett,avgoustidis2012} and models in which a cosmological scalar
field couples directly to the electromagnetic sector
\cite{carroll,hees}. The near-blackbody CMB spectrum and direct
measurements of the CMB temperature at non-zero redshift therefore provide
complementary tests of non-standard photon evolution
\cite{fixsen1996,lima2,luzzi,hurier}.

Scalar--photon interactions are particularly interesting because the photon
thermal history, recombination and acoustic perturbations are then linked to
the dynamics of the scalar field rather than being modified independently.
Previous treatments of scalar--radiation energy exchange and modified
temperature--redshift relations have often used phenomenological source terms
or approximately constant deviations from the standard photon dilution law
\cite{lima_vac,avgoustidis2012}. In the present work we instead consider
a canonical scalar field coupled exclusively to the cosmological
photon-fluid Lagrangian through an exponential factor. Baryons, cold dark matter (CDM), neutrinos and the late-time
vacuum component remain uncoupled while the scalar field evolves in an
exponential potential. This setup is also motivated by interacting
early-time scalar field scenarios in which energy exchange with radiation
affects the pre-recombination evolution \cite{bis2}.

The coupled radiation era system admits a scaling branch on which the departure
from the standard photon dilution law is constant \cite{bis1}. We characterize the branch
used to initialize the numerical solution by an early-time parameter
$\epsilon_i$ while $\sigma$ denotes the scalar--photon coupling strength.
These two quantities fix the corresponding exponential potential slope.
Crucially, the constant scaling relation is used only to select the regular
radiation era initial trajectory. After initialization, the complete coupled
background equations are integrated and the instantaneous departure from
standard photon dilution evolves dynamically. Thus the present model is not a
cosmology in which a constant $\epsilon$ is imposed at all epochs.

The principal aim of this work is to connect the effective interaction to
observable cosmology. We derive the background and linear scalar perturbation
equations in synchronous gauge, including leading-order adiabatic initial relations,
and implement the system in CLASS \cite{class}. We propagate the evolving
photon temperature, under an explicit thermal-bath assumption, into the
recombination calculation and examine the expansion history, visibility
function, CMB temperature and polarization spectra, lensing, and the linear
matter power spectrum. The implementation is tested against the matched
interaction-off $\Lambda$CDM solution and archived precision-setting checks.
We additionally check the scalar and photon fluid-level coupling terms and
numerically test a correction to the tight-coupling derivative; at the tested
weakly interacting point, its maximum relative effect on the TT spectrum
through $\ell=2500$ is approximately $2.4\times10^{-6}$. These checks do
not establish independence from the numerical initialization epoch or a
microscopic mechanism for maintaining a Planckian photon distribution, and
they do not constitute a derivation of the full interacting photon Boltzmann
hierarchy.

We distinguish two uses of the numerical model. First, a comparatively large
fiducial interaction, $(\epsilon_i,\sigma)=(0.023,0.05)$, is used to expose
its physical response: shifts in the visibility maximum and acoustic scale,
oscillatory CMB residuals, and scale-dependent matter-spectrum changes. These
are illustrative predictions, not a best-fitting cosmology. Second, we
calculate the Planck high-$\ell$ TTTEEE-lite, low-$\ell$ TT and EE, and
CMB-lensing likelihoods on a conditional grid in $(\epsilon_i,\sigma)$.
The six standard cosmological parameters are held at the adopted baseline,
while $A_{\rm Planck}$ is profiled. This exercise measures sensitivity at
that baseline; it is not a joint parameter inference and cannot establish
marginalized bounds or the effect of cosmological-parameter degeneracies.

The paper is organized as follows. Section~2 presents the model, radiation-era
initial conditions, perturbations, and the thermodynamic assumptions.
Section~3 describes the CLASS implementation and its predictions for
recombination and cosmological observables. Section~4 reports the conditional
Planck likelihood comparison, including the separately evaluated
interaction-off reference. Section~5 discusses the astrophysical implications,
statistical limitations, and the physical consistency tests required for
further constraints. The appendices give details of the perturbation sources,
adiabatic initial conditions, and supplementary CMB responses.


\section{Scalar--photon coupling: background and perturbations}
\subsection{Action and background equations}
We consider a canonical scalar field $\phi$ coupled to the cosmological
photon fluid while baryons, CDM, neutrinos and the late-time vacuum
component remain uncoupled. The effective action is\footnote{We set $8\pi G=c=\hbar=1$.}
\begin{equation}
S =
\int d^4x\,\sqrt{-g}
\left[
\frac{1}{2}R
-\frac{1}{2}g^{\mu\nu}\nabla_\mu\phi\nabla_\nu\phi
-V(\phi)
+e^{-\sigma\phi}\mathcal{L}_{\gamma}
+\mathcal{L}_{u}
\right],
\label{eq:action}
\end{equation}
where $V(\phi)$ is the potential, $\sigma$ is
a constant coupling parameter, $\mathcal{L}_{\gamma}$ denotes the
Lagrangian density of the cosmological photon fluid and
$\mathcal{L}_{u}$ contains the uncoupled components.
Throughout this work, $\mathcal{L}_{\gamma}$ is understood as an effective
perfect fluid description of the thermal photon bath rather than as a
scalar dependent modification of the microscopic Maxwell kinetic term.
We adopt the on-shell perfect fluid representation [27, 28, 29]
\begin{equation}
\mathcal{L}_{\gamma}=p_{\gamma}=\frac{1}{3}\rho_{\gamma},
\label{eq:photon_lagrangian}
\end{equation}
The interaction in
Eq.~\eqref{eq:action} should therefore be regarded as an effective
description of energy-momentum exchange between the scalar field and
the cosmological photon fluid. In particular, the present framework does
not include microscopic electromagnetic effects such as a
scalar dependent fine structure constant or corresponding modifications
of atomic transition rates.

Variation of Eq.~\eqref{eq:action} with respect to the metric and the
scalar field gives, respectively,
\begin{equation}
G_{\mu\nu}
=
T_{\mu\nu}^{(\phi)}
+
e^{-\sigma\phi}T_{\mu\nu}^{(\gamma)}
+
T_{\mu\nu}^{(u)},
\label{eq:einstein}
\end{equation}
and
\begin{equation}
\nabla_{\mu}\nabla^{\mu}\phi
-
V_{,\phi}
=
\sigma e^{-\sigma\phi}\mathcal{L}_{\gamma},
\label{eq:scalar_covariant}
\end{equation}
where
\begin{equation}
T^{(\phi)}_{\mu\nu}
=
\nabla_{\mu}\phi\nabla_{\nu}\phi
-
\frac{1}{2}g_{\mu\nu}(\nabla\phi)^2
-
g_{\mu\nu}V(\phi),
\end{equation}
and $T^{(\gamma)}_{\mu\nu}$ and $T^{(u)}_{\mu\nu}$ are energy-momentum tensors corresponding to photons and the uncoupled components, respectively. 
Using the scalar field equation, the divergence of
$T_{\mu\nu}^{(\phi)}$ is
\begin{equation}
\nabla^\mu T_{\mu\nu}^{(\phi)}
=
Q_\nu ,
\label{eq:scalar_transfer}
\end{equation}
where
\begin{equation}
Q_\nu
\equiv
\sigma e^{-\sigma\phi}\mathcal{L}_{\gamma}\nabla_\nu\phi
=
\frac{\sigma}{3}
e^{-\sigma\phi}\rho_{\gamma}\nabla_\nu\phi .
\label{eq:Q_def}
\end{equation}
The Bianchi identities, $\nabla^\mu G_{\mu\nu}=0$, then require
\begin{equation}
\nabla^\mu
\left(
e^{-\sigma\phi}T_{\mu\nu}^{(\gamma)}
\right)
=
-Q_\nu .
\label{eq:weighted_photon_conservation}
\end{equation}
Thus the scalar and photon sectors are not separately conserved and they
exchange energy--momentum while their combined contribution is
covariantly conserved.

For a spatially flat FLRW background, the field equations become
\begin{equation}
3H^2
=e^{-\sigma\phi}\rho_\gamma+\rho_\phi+\rho_u,
\label{eq:full-friedmann}
\end{equation}
\begin{equation}
\ddot\phi+3H\dot\phi+V_{,\phi}
=-\frac{\sigma}{3}e^{-\sigma\phi}\rho_\gamma,
\label{a2}
\end{equation}
\begin{equation}
\dot\rho_\gamma+4H\rho_\gamma
=\frac{4\sigma}{3}\dot\phi\,\rho_\gamma,
\label{a3}
\end{equation}
where
\begin{equation}
\rho_\phi=\frac{1}{2}\dot\phi^2+V(\phi).
\end{equation}
Equation~(\ref{a3}) may be integrated exactly as
\begin{equation}
\rho_\gamma(a)
=
\rho_{\gamma0}\,a^{-4}
\exp\!\left[
\frac{4\sigma}{3}
\left(\phi(a)-\phi_0\right)
\right],
\label{a5}
\end{equation}
where $a_0=1$, $\rho_{\gamma0}\equiv\rho_\gamma(a=1)$ is the present
photon-fluid density and $\phi_0\equiv\phi(a=1)$. The standard photon-dilution law is recovered
when the scalar--photon energy exchange is switched off.

\paragraph{Thermal and gravitational photon densities.}
The photon density $\rho_\gamma$ appearing in the effective fluid
Lagrangian must be distinguished from the density sourcing the Einstein
equations, $\widetilde\rho_\gamma\equiv e^{-\sigma\phi}\rho_\gamma$.
Combining this definition with equation~\eqref{a3} yields
\begin{equation}
\dot{\widetilde\rho}_\gamma+4H\widetilde\rho_\gamma
=\frac{\sigma}{3}\dot\phi\,\widetilde\rho_\gamma.
\label{eq:weighted-photon-background}
\end{equation}
In particular, $\widetilde\rho_{\gamma0}=e^{-\sigma\phi_0}\rho_{\gamma0}$.
The measured present CMB temperature fixes $\rho_{\gamma0}$ through the
blackbody relation only if $\rho_\gamma$ is identified with the physical
thermal energy density. It does not separately fix the coupling
normalization $e^{-\sigma\phi_0}$. The latter must be specified by the
model's normalization convention or fixed by an additional physical
condition; it cannot be inferred from $T_0$ alone. This distinction is
important when comparing the interacting model with a standard CLASS
radiation density and when assessing the shooting calibration.

\subsection{Radiation era scaling branch and dynamical photon temperature}

A useful measure of the instantaneous departure from standard photon dilution
is
\begin{equation}
\epsilon(a)
\equiv
\frac{d\ln\rho_\gamma}{d\ln a}+4
=
\frac{4\sigma}{3}\frac{d\phi}{d\ln a}.
\label{a7}
\end{equation}
In general, $\epsilon(a)$ is a dynamical quantity. The coupled system,
however, admits a radiation era scaling branch for which $\epsilon(a)$ is
constant \cite{bis2}. We denote the value characterizing this early-time branch by
$\epsilon_i$. On this solution,
\begin{equation}
\frac{d\phi}{d\ln a}
=
\frac{3\epsilon_i}{4\sigma},
\label{eq:phi-scaling}
\end{equation}
and hence, relative to an initial epoch $a_i$ within the scaling regime,
\begin{equation}
\rho_\gamma(a)
=
\rho_{\gamma i}
\left(\frac{a}{a_i}\right)^{-4+\epsilon_i}.
\label{b3}
\end{equation}
Thus $\epsilon_i$ specifies the early-time deviation from the usual
$a^{-4}$ photon dilution law.

Assuming that the cosmological photon bath remains thermal with negligible
chemical potential, its energy density satisfies
$\rho_\gamma\propto T_\gamma^4$. Equation~(\ref{a5}) therefore
implies the general temperature evolution
\begin{equation}
T_\gamma(a)
=
T_0\,a^{-1}
\exp\!\left[
\frac{\sigma}{3}
\left(\phi(a)-\phi_0\right)
\right],
\label{eq:dynamic-temperature}
\end{equation}
or equivalently
\begin{equation}
T_\gamma(z)
=
T_0(1+z)
\exp\!\left[
\frac{\sigma}{3}
\left(\phi(z)-\phi_0\right)
\right].
\label{eq:dynamic-temperature-redshift}
\end{equation}
On the radiation era scaling branch this reduces locally to
\begin{equation}
T_\gamma(a)
=
T_{\gamma i}
\left(\frac{a}{a_i}\right)^{-1+\epsilon_i/4}.
\label{a14}
\end{equation}
\paragraph{Thermalization assumption and spectral distortions.}
The temperature relations above are conditional on the assumption that
the interacting photon bath remains Planckian. The effective fluid action
and its energy-continuity equation do not determine the photon phase-space
distribution or the photon-number evolution. Consequently, they do not by
themselves demonstrate that energy exchange preserves a blackbody spectrum.
In particular, energy transfer after efficient photon-number-changing
thermalization has ceased may generate chemical-potential or Compton-$y$
distortions, subject to CMB spectral measurements. A microscopic kinetic
prescription, or an independently justified thermalization mechanism, is
needed to predict such distortions and to turn the background energy
transfer into a spectral-distortion bound. In particular, the Planck
anisotropy likelihood evaluated below does not constitute a test of the
COBE/FIRAS blackbody-spectrum constraint \cite{fixsen1996}; no FIRAS
likelihood or spectral-distortion limit is imposed in this analysis.
The present temperature law should therefore be interpreted as a
phenomenological thermal-bath closure, not as a derived prediction for
an arbitrary non-equilibrium photon distribution. The dimensionless accumulated modification
\begin{equation}
\Delta_\gamma(a_1,a_2)\equiv
\ln\!\left[\frac{\rho_\gamma(a_2)a_2^4}
{\rho_\gamma(a_1)a_1^4}\right]
=\int_{\ln a_1}^{\ln a_2}\epsilon(a)\,\mathrm{d}\ln a
\label{eq:cumulative-photon-exchange}
\end{equation}
quantifies the departure from standard dilution but is not itself a
spectral-distortion observable. Quantitative compatibility with spectral
constraints remains to be demonstrated for the phenomenological closure
used here.

The power-law form in equation~(\ref{a14}) is therefore an early-time scaling
relation rather than a temperature--redshift law imposed at all epochs. Once
the numerical evolution leaves the initial scaling regime,
$\phi(a)$ is obtained from the complete background system and
$T_\gamma(a)$ follows dynamically from equation~(\ref{eq:dynamic-temperature}).

We close the scalar sector with an exponential potential,
\begin{equation}
V(\phi)=V_0e^{-\lambda\phi}.
\label{eq:exponential-potential}
\end{equation}
On the radiation era scaling branch,
\begin{equation}
V\propto
a^{-3\lambda\epsilon_i/(4\sigma)},
\end{equation}
whereas the photon contribution to the Einstein equations scales as
\begin{equation}
e^{-\sigma\phi}\rho_\gamma
\propto a^{-4+\epsilon_i/4}.
\end{equation}
Requiring the scalar potential and the photon contribution to the Einstein
equations to scale with the same power of $a$ on the radiation era scaling
branch gives
\begin{equation}
\lambda
=
\frac{\sigma(16-\epsilon_i)}{3\epsilon_i}.
\label{eq:lambda-epsilon-sigma}
\end{equation}
Consequently, $\epsilon_i$ and $\sigma$ determine the radiation era scaling
trajectory and fix the potential slope $\lambda$ which is not treated as an
independent parameter. After initialization, $\lambda$ remains fixed while
the full background equations determine $\phi(a)$ and hence the instantaneous
quantity $\epsilon(a)$.

The radiation era scaling parametrization is defined on the interacting
branch. Equation~\eqref{eq:phi-scaling} assumes $\sigma\neq0$ while equation~\eqref{eq:lambda-epsilon-sigma} becomes
singular as $\epsilon_i\rightarrow0$. The interaction-off cosmology used
later as the $\Lambda$CDM reference is therefore implemented as a separate
branch in which the scalar--photon interaction is switched off rather than
by identifying $\epsilon_i=0$ as a regular point of the interacting scaling
parametrization.
\subsection{Linear scalar perturbations}

We use the synchronous-gauge conventions of CLASS \cite{class},
\begin{equation}
ds^2=a^2(\tau)\left[
-d\tau^2+
(\delta_{ij}+h_{ij})dx^idx^j
\right],
\qquad
h_{ij}
=
\hat k_i\hat k_j h
+6\eta
\left(
\hat k_i\hat k_j-\frac{1}{3}\delta_{ij}
\right),
\label{eq:synchronous-metric}
\end{equation}
with ${\cal H}=a'/a$. Writing
\begin{equation}
\phi=\bar\phi+\delta\phi,
\qquad
\rho_\gamma
=
\bar\rho_\gamma(1+\delta_\gamma),
\end{equation}
the scalar perturbation obeys
\begin{align}
\delta\phi''
+2{\cal H}\delta\phi'
+\frac{1}{2}h'\bar\phi'
+\left[
k^2+a^2V_{,\phi\phi}
-\frac{a^2\sigma^2}{3}
e^{-\sigma\bar\phi}\bar\rho_\gamma
\right]\delta\phi
={}&
-\frac{a^2\sigma}{3}
e^{-\sigma\bar\phi}\bar\rho_\gamma
\delta_\gamma.
\label{eq:scalar-perturbation}
\end{align}

Because the photon stress-energy tensor appears in the Einstein equations
multiplied by $e^{-\sigma\phi}$, its density perturbation contributes through
\begin{equation}
\delta\!\left(
e^{-\sigma\phi}\rho_\gamma
\right)
=
e^{-\sigma\bar\phi}\bar\rho_\gamma
\left(
\delta_\gamma-\sigma\delta\phi
\right).
\label{eq:weighted-photon-density-perturbation}
\end{equation}
The combination $\delta_\gamma-\sigma\delta\phi$ therefore enters the
perturbed Einstein sources but we do not regard it as the density contrast
of an additional fluid. The remaining photon contributions to the perturbed
stress-energy tensor are given in
appendix~\ref{app:perturbation-sources}.

For the isotropic energy transfer closure adopted in the numerical
implementation, the photon monopole equation becomes
\begin{equation}
\delta_\gamma'
+\frac{4}{3}\theta_\gamma
+\frac{2}{3}h'
=
\frac{4\sigma}{3}\delta\phi',
\label{eq:photon-continuity-coupled}
\end{equation}
whereas the Euler equation retains its standard form,
\begin{equation}
\theta_\gamma'
=
k^2
\left(
\frac{\delta_\gamma}{4}-\Pi_\gamma
\right)
+a n_e\sigma_{\rm T}
(\theta_b-\theta_\gamma).
\label{eq:photon-euler-coupled}
\end{equation}
The higher photon temperature and polarization multipoles as well as the
uncoupled baryon, CDM and neutrino equations retain their
standard CLASS structure. Their solutions are nevertheless modified through
the altered background evolution, the photon monopole and the Einstein
sources generated by the coupled scalar--photon system.

\subsection{Adiabatic initial conditions}

The perturbations are initialized deep in the radiation era with
$k\tau_i\ll1$ and
$a n_e\sigma_{\rm T}\gg{\cal H},k$. At this time the background is placed on
the scaling branch defined above so that
\begin{equation}
\bar\phi'
=
\frac{3\epsilon_i}{4\sigma}{\cal H}.
\label{eq:initial-phi-prime}
\end{equation}
For the regular growing adiabatic mode, all species correspond to the same
local time shift. Since the photon density scales initially as
$a^{-4+\epsilon_i}$, the density contrasts satisfy
\begin{equation}
\frac{\delta_\gamma}{4-\epsilon_i}
=
\frac{\delta_\nu}{4}
=
\frac{\delta_b}{3}
=
\frac{\delta_c}{3}.
\label{eq:adiabatic-density-relations}
\end{equation}
For the leading synchronous-gauge mode
\begin{equation}
h=C(k\tau)^2,
\end{equation}
the density perturbations are
\begin{equation}
\delta_{\gamma i}
=
-\frac{4-\epsilon_i}{6}h_i,
\qquad
\delta_{\nu i}
=
-\frac{2}{3}h_i,
\qquad
\delta_{b i}
=
\delta_{c i}
=
-\frac{1}{2}h_i.
\label{eq:initial-density-conditions}
\end{equation}
The corresponding scalar field perturbations are
\begin{equation}
\delta\phi_i
=
\frac{\epsilon_i}{8\sigma}h_i,
\qquad
\delta\phi_i'
=
\frac{\epsilon_i}{8\sigma}h_i'.
\label{eq:initial-scalar-conditions}
\end{equation}
The derivation, including the consistency of the photon contribution to the
perturbed Einstein equations, is given in
appendix~\ref{app:adiabatic-initial-conditions}. The residual
synchronous-gauge freedom is fixed in the CDM rest frame,
$\theta_{c i}=0$ while tight coupling imposes
$\theta_{\gamma i}=\theta_{b i}$. For the remaining species, we use the standard CLASS adiabatic coefficients
at the order implemented. The relations above follow from the common
time-shift prescription and the leading photon continuity equation; they do
not constitute an independent higher-order super-Hubble eigenmode analysis
of the complete interacting Einstein--Boltzmann system. When the interaction
is disabled, the additional scalar--photon contributions vanish and the
usual CLASS initial vector is recovered.

\section{Numerical implementation, validation and physical signatures}
\label{sec:numerical-results}

The background and perturbation system derived in Section~2 was implemented
in the Boltzmann code CLASS. The radiation era scaling solution is used only
to set the initial background and perturbation variables. Thereafter, the
complete coupled system is evolved numerically so that the scalar field, the
photon dilution rate and the photon temperature evolve dynamically. The
resulting $T_\gamma(z)$ is supplied consistently to the recombination
calculation while the scalar--photon perturbation terms and the modified
adiabatic initial conditions are included in the Einstein--Boltzmann system.

In this section we describe the numerical implementation and its validation
and illustrate the physical response of the model using parameter values
chosen to make the effects visible. These calculations are controlled
fixed-parameter response tests and are distinct from the conditional
profile-likelihood analysis
presented in Section~4.

\subsection{CLASS implementation and validation}
\label{subsec:class-implementation-validation}

The interacting model is specified by the scalar--photon coupling strength
$\sigma$ and by the radiation era scaling parameter $\epsilon_i$. For each
pair $(\epsilon_i,\sigma)$, the exponential potential slope is fixed by
equation~\eqref{eq:lambda-epsilon-sigma} and is therefore not an independent parameter. The background
is initialized on the corresponding radiation era scaling trajectory and is
subsequently evolved by integrating the full coupled equations. In
particular, neither a constant value of $\epsilon$ nor a power-law
temperature--redshift relation is imposed after initialization.

The normalization of the initial photon density, the normalization of the
scalar potential and the late-time vacuum contribution are determined by an
external shooting procedure. These quantities are adjusted so that the
specified present photon density and Hubble rate are recovered while the
initial scalar trajectory satisfies the required scaling conditions. They
are therefore calibration variables selecting the desired cosmological
trajectory rather than independently sampled interaction parameters. The shooting procedure is repeated whenever the interaction parameters
or the baseline cosmological parameters entering these conditions are changed.
The background shooting procedure calibrates the bare photon density
$\rho_{\gamma0}$ to the value inferred from the adopted present-day CMB
temperature, while the photon contribution to the Einstein equations is the
weighted density
$\widetilde\rho_{\gamma0}=e^{-\sigma\phi_0}\rho_{\gamma0}$.
The shooting parameters are adjusted simultaneously to match the target
Hubble parameter and the initial radiation-era scaling condition. For the
representative interacting solution
$(\epsilon_i,\sigma)=(4\times10^{-5},0.10)$, the residuals in the
photon-density and Hubble-parameter targets are $1.93\times10^{-9}$ and
zero at the reported precision, respectively. This calibration is
internally consistent with the effective-fluid implementation; the
identification of the bare density with the observed thermal photon density
is a phenomenological assumption rather than a consequence of a microscopic
photon kinetic theory. The shooting diagnostics do not independently fix
the physical coupling normalization $e^{-\sigma\phi_0}$ or establish
independence from the chosen initialization epoch.

The implementation modifies the CLASS background, thermodynamics and
perturbation modules. In the background sector, the scalar field and photon
density are evolved simultaneously and the photon contribution to the
gravitational equations is included with its explicit
$e^{-\sigma\phi}$ weighting. The dynamically evolved photon temperature is
passed consistently to the recombination module. The production likelihood
calculations reported below use RecFast while the corresponding HyRec
interface was used as an independent thermodynamic cross-check where the
evolved temperature remains within its supported range. In the perturbation
sector, the scalar fluctuation, the modified photon monopole and the
corresponding photon contributions to the Einstein sources are evolved
together with the initial conditions~\eqref{eq:initial-density-conditions}--\eqref{eq:initial-scalar-conditions}. The higher photon
temperature and polarization hierarchies retain their standard CLASS
structure. The usual CLASS output modules are then used to compute unlensed
and lensed CMB spectra, the lensing-potential spectrum and the linear matter
power spectrum.

The implementation was tested independently at the background,
thermodynamic and perturbation levels. During the initial
radiation-dominated regime, the numerical photon evolution reproduces the
analytic scaling solution. All background, thermodynamic and perturbation
variables remain finite throughout the physical integration interval and
switching off the interaction recovers the matched $\Lambda$CDM background
and spectra. The resulting unlensed and lensed spectra remain numerically
stable through $\ell=3000$.

For the conditional profile-likelihood calculations of Section~4, the
shooting procedure uses a hard acceptance tolerance of $10^{-6}$ for the
largest calibration residual while a substantially tighter target of
$5\times10^{-8}$ is imposed on the final exact profile calculations. All
interacting cosmologies retained in the final profile analysis satisfy this
tighter target. The largest residual in that set is approximately
$4.14\times10^{-8}$. The archived exploratory runs used the same calibrated
background construction at each sampled cosmology, with failed calibrations
rejected before likelihood evaluation. The tighter $5\times10^{-8}$
refinement quoted above applies specifically to the final exact profile
points.
\par\medskip
\begin{center}
\begin{tabular}{ll}
\hline
Test & Result \\
\hline
Radiation era scaling &
Numerical photon evolution reproduces the analytic branch \\
Background calibration &
Hard acceptance tolerance $10^{-6}$ \\
Profile-point precision &
Final exact points satisfy $r_{\rm max}<5\times10^{-8}$ \\
Exploratory calibration &
Failed background calibration trials are rejected \\
Thermodynamics &
RecFast production; HyRec cross-checks within its supported range \\
Perturbations and initial conditions &
Coupled system finite and regular \\
Matched $\Lambda$CDM regression &
Interaction-off background and spectra recovered \\
Spectral stability &
Stable unlensed and lensed spectra through $\ell=3000$ \\
\hline
\end{tabular}
\end{center}
\refstepcounter{table}
\label{tab:class-validation}
\noindent{\small\textbf{Table~\thetable:}
Summary of the numerical validation of the scalar--photon implementation.
The calibration residual is the maximum of the photon-normalization,
Hubble rate and initial-trajectory residuals. The hard shooting acceptance
criterion is $10^{-6}$ while the exact cosmologies entering the final
profile-likelihood analysis were refined to a target below $5\times10^{-8}$.
The largest residual among the retained final points is
$4.14\times10^{-8}$.\par}
\par\medskip

\subsection{Background evolution and recombination}
\label{subsec:background-recombination}

To display the physical effects of the interaction, we consider
$\epsilon_i=0.015$, $0.023$ and $0.030$ at fixed $\sigma=0.05$ and use
\begin{equation}
(\epsilon_i,\sigma)=(0.023,0.05)
\label{eq:fiducial-illustrative-point}
\end{equation}
as an illustrative fiducial point. These relatively large values are chosen
to make the physical response clearly visible and lie far from the
small-$\epsilon_i$ region relevant to the Planck likelihood analysis.
They should therefore not be interpreted as observationally preferred
parameters.

For every value of $\epsilon_i$, the auxiliary trajectory variables are
recalibrated independently so that the same present photon density and
Hubble rate targets are recovered. The ordinary cosmological parameters are
held fixed in these comparisons. The curves therefore isolate the physical
response to the scalar--photon interaction and are not intended to follow
the cosmological-parameter degeneracy directions explored in the
conditional analysis of Section~4. The matched $\Lambda$CDM reference is
obtained by switching off the scalar--photon interaction while retaining the
same ordinary cosmological parameters and present-day targets.

On the initial radiation era scaling branch, positive $\epsilon_i$ implies a
slower dilution of the unweighted photon density than $a^{-4}$. For a fixed
present temperature, the photon temperature is correspondingly reduced
relative to the standard value when evolved back to sufficiently high
redshift. After initialization, however, the temperature is not obtained from
a fixed power law, it follows the numerical scalar trajectory through
equation~(\ref{eq:dynamic-temperature-redshift}).

\begin{figure}[!htbp]
    \centering
    \includegraphics[width=0.94\textwidth]{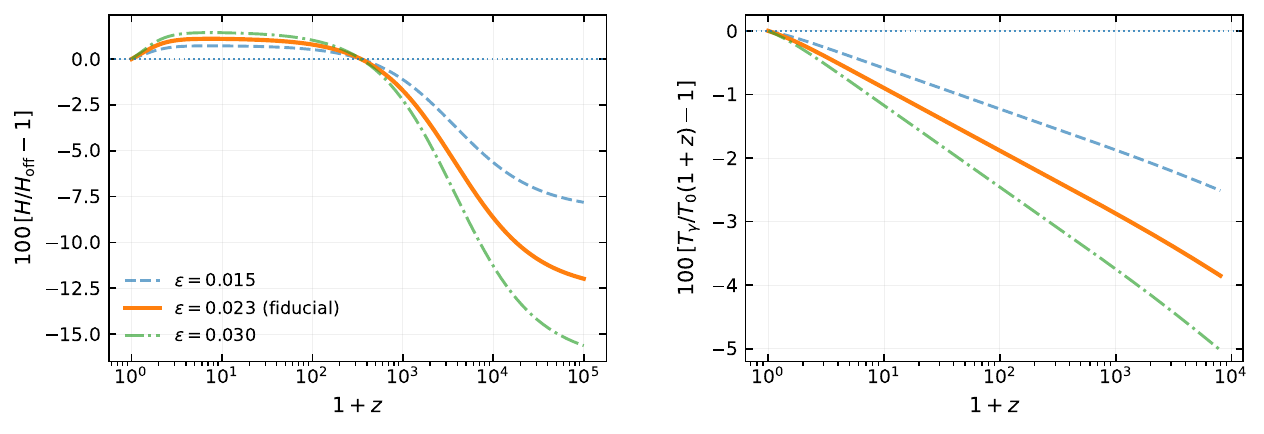}
    \caption{Illustrative fixed-parameter response of the background expansion
rate and photon temperature at fixed $\sigma=0.05$ for
$\epsilon_i=0.015$, $0.023$ and $0.030$. Each interacting solution is
    independently calibrated to the same present photon density and
    Hubble rate targets. The solid curve denotes the illustrative fiducial
    value $\epsilon_i=0.023$. The left panel shows
    $100[H/H_{\Lambda{\rm CDM}}-1]$ relative to the matched interaction-off
    $\Lambda$CDM model. The right panel shows
    $100[T_\gamma/T_0(1+z)-1]$ where $T_\gamma(z)$ is obtained from the full
    dynamical scalar--photon evolution rather than from an imposed constant
    power law. The horizontal dotted lines mark zero deviation.}
    \label{fig:background-temperature}
\end{figure}

The altered photon contribution and the calibrated scalar trajectory jointly
modify the expansion rate. Figure~\ref{fig:background-temperature} shows the
fractional changes in $H(z)$ and $T_\gamma(z)$ for the three illustrative
models. The sign change in the Hubble rate response reflects the competition
between the modified photon contribution and the scalar and late-time
components required by the common present-day calibration. Each curve
therefore represents a complete independently calibrated solution rather than
a rescaling of the fiducial model.

\begin{figure}[!htbp]
    \centering
    \includegraphics[width=0.85\textwidth]{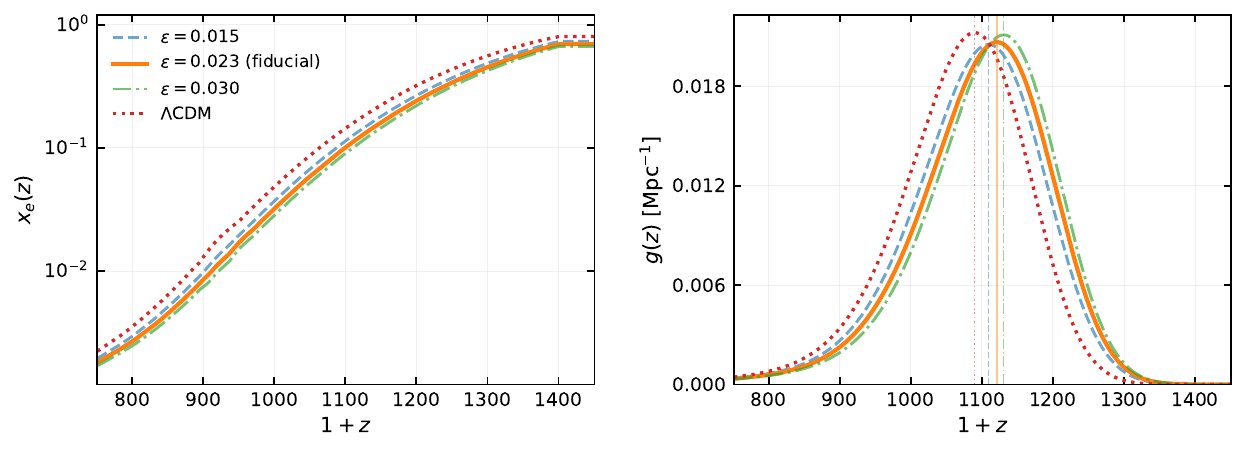}
   \caption{
Illustrative fixed-parameter response of the ionization history and visibility
function at fixed $\sigma=0.05$ for $\epsilon_i=0.015$, $0.023$ and
$0.030$ together with the matched interaction-off $\Lambda$CDM reference.
Each interacting model is independently calibrated to the same present-day
photon-density and Hubble rate targets while the ordinary cosmological
parameters are held fixed. The solid curve denotes the fiducial value
$\epsilon_i=0.023$. The left panel shows the free-electron fraction
$x_e(z)$ and the right panel the visibility function $g(z)$. The vertical
lines mark the corresponding visibility maxima and show the displacement of
the last scattering surface produced by the scalar--photon interaction.
These parameter choices are selected to make the physical response visible
and are not observationally preferred values.
}
    \label{fig:recombination-visibility}
\end{figure}

The dynamically evolved photon temperature is propagated directly into the
recombination calculation. Figure~\ref{fig:recombination-visibility} shows the
free-electron fraction
\begin{equation}
x_e(z)\equiv\frac{n_e(z)}{n_{\rm H}(z)}
\label{eq:xe-definition-section3}
\end{equation}
and the visibility function $g(z)$. We define the last-scattering redshift by
\begin{equation}
z_*\equiv\operatorname*{arg\,max}_z g(z)
\end{equation}
and its displacement relative to the matched reference by
\begin{equation}
\Delta z_*
\equiv
z_*^{\rm fid}-z_*^{\Lambda{\rm CDM}}.
\end{equation}
For the illustrative fiducial model,
\begin{equation}
\Delta z_*\simeq+31.13,
\end{equation}
so that the maximum of the visibility function moves to higher redshift.
Denoting the full width at half maximum of the visibility peak by $w_g$, its
absolute fractional change is
\begin{equation}
\frac{|\Delta w_g|}
     {w_g^{\Lambda{\rm CDM}}}
\simeq2.59\%.
\end{equation}

The acoustic angular scale is
\begin{equation}
\theta_*
\equiv
\frac{r_s(z_*)}{D_M(z_*)},
\end{equation}
where $r_s(z_*)$ is the comoving sound horizon and $D_M(z_*)$ is the
transverse comoving distance. For the same fiducial model,
\begin{equation}
\frac{|\Delta\theta_*|}
     {\theta_*^{\Lambda{\rm CDM}}}
\simeq3.41\%.
\end{equation}
The modified thermal and expansion histories therefore alter both the location
and the width of the last-scattering surface and shift the acoustic scale,
providing the main physical origin of the CMB response discussed below.
These values refer to an illustrative fixed-parameter model, not to the
best sampled point in the conditional Planck analysis. In particular, a
change in $\theta_*$ alone does not demonstrate an observationally viable
reduction in the sound horizon or a resolution of the Hubble tension;
such a conclusion would require a joint parameter fit including appropriate
late-time distance information.

\subsection{CMB and matter power spectra}
\label{subsec:cmb-matter-spectra}

The complete perturbation evolution propagates the scalar--photon interaction
into the temperature, polarization and lensing spectra. For an auto-spectrum
$XX$, we define the fiducial percentage response
\begin{equation}
{\cal R}_{\ell}^{XX}
\equiv
100\left(
\frac{C_{\ell,\mathrm{fid}}^{XX}}
     {C_{\ell,\Lambda{\rm CDM}}^{XX}}-1
\right),
\qquad
XX\in\{TT,EE,BB,\phi\phi\},
\label{eq:auto-spectrum-response}
\end{equation}
where ``fid'' denotes the illustrative interacting model defined in
equation~\eqref{eq:fiducial-illustrative-point}, and the denominator is the
matched interaction-off $\Lambda$CDM spectrum. Since cross-spectra can change sign or
pass through zero, we instead define
\begin{equation}
{\cal D}_{\ell}^{XY}
\equiv
100\,
\frac{
C_{\ell,\mathrm{fid}}^{XY}
-C_{\ell,\Lambda{\rm CDM}}^{XY}
}{
\sqrt{
C_{\ell,\Lambda{\rm CDM}}^{XX}
C_{\ell,\Lambda{\rm CDM}}^{YY}
}},
\qquad
XY\in\{TE,T\phi,E\phi\}.
\label{eq:cross-spectrum-response}
\end{equation}

\begin{figure}[!htbp]
    \centering
    \includegraphics[width=\textwidth]{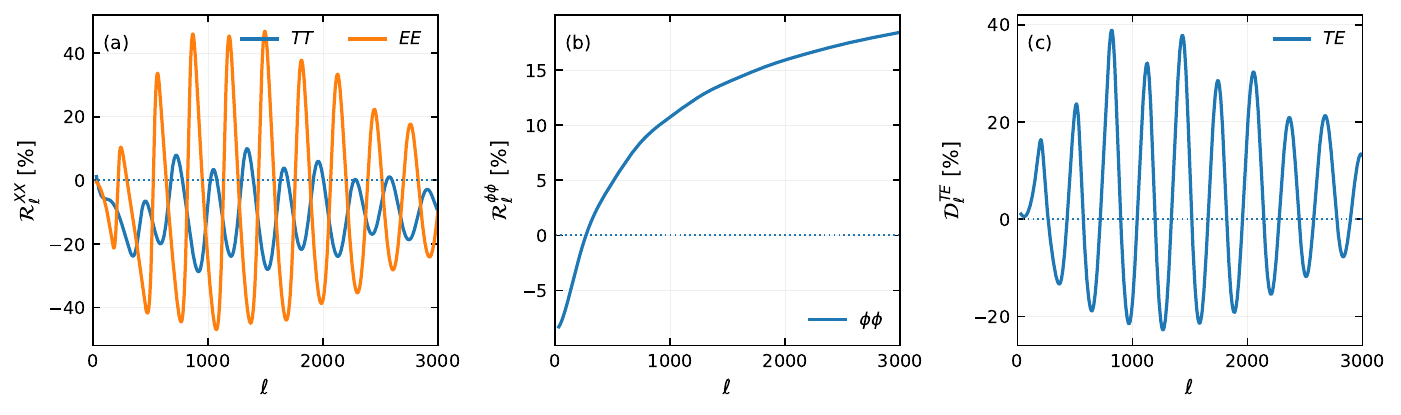}
    \caption{Principal lensed-CMB responses of the illustrative fiducial
    scalar--photon model,
    $(\epsilon_i,\sigma)=(0.023,0.05)$ relative to the matched
    interaction-off $\Lambda$CDM reference over
    $2\leq\ell\leq3000$. Panel (a) shows the percentage TT and EE
    auto-spectrum responses ${\cal R}_{\ell}^{XX}$, panel (b) the
    lensing-potential response ${\cal R}_{\ell}^{\phi\phi}$ and panel (c)
    the normalized TE difference ${\cal D}_{\ell}^{TE}$. The oscillatory
    residuals mainly trace the coherent displacement of the acoustic pattern
    with changes in diffusion damping and acoustic driving modifying their
    envelope. Large local percentage values therefore do not correspond to a
    uniform amplitude rescaling.}
    \label{fig:lensed-cmb-residuals}
\end{figure}

Figure~\ref{fig:lensed-cmb-residuals} shows the principal CMB signatures over
$2\leq\ell\leq3000$. The TT, EE and TE residuals arise from several coupled
effects. The change in $\theta_*$ found in
Subsection~\ref{subsec:background-recombination} coherently shifts the
acoustic pattern and produces the dominant alternating residuals. The
modified recombination history and visibility width alter diffusion damping,
particularly toward high multipoles while the changed background and metric
perturbations modify acoustic driving and peak amplitudes.

The comparatively large local percentage extrema, especially in EE, should
therefore not be interpreted as a uniform change in spectral amplitude. They
occur predominantly because the interacting and reference spectra are
evaluated on opposite sides of displaced acoustic extrema or because the
reference spectrum is locally small. By contrast, the response of the
lensing-potential spectrum is smoother and reflects the integrated change in
the gravitational potentials along the line of sight. For clarity, the
lower-priority BB, $T\phi$ and $E\phi$ responses are shown in
appendix~\ref{app:supplementary-cmb-responses}.

\begin{figure}[!htbp]
    \centering
    \includegraphics[width=0.82\textwidth]
    {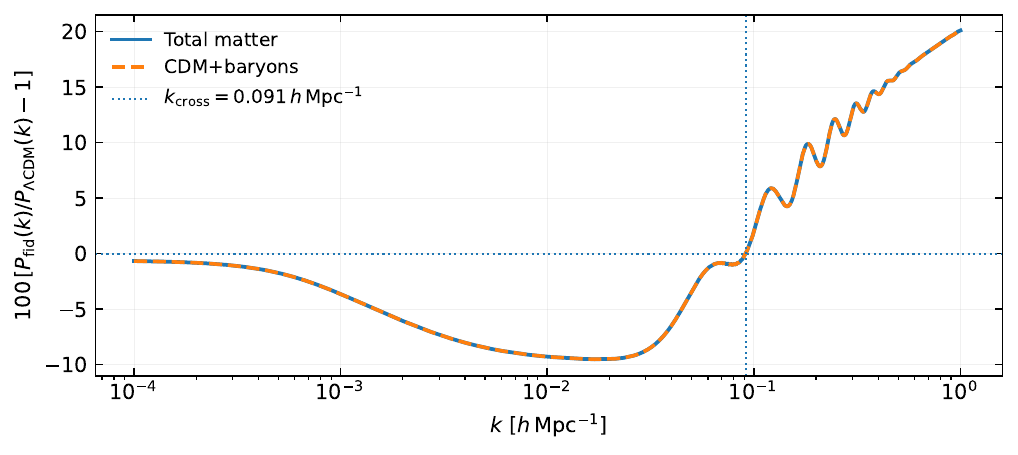}
    \caption{Scale dependent fractional response of the linear matter power
    spectrum at $z=0$ for the illustrative fiducial model
    $(\epsilon_i,\sigma)=(0.023,0.05)$ relative to the matched interaction-off
    $\Lambda$CDM reference. The solid and dashed curves denote the total matter
    and CDM-plus-baryon responses, respectively. Their near overlap shows that
    the reshaping is already present in the clustered matter sector. The
    oscillatory modulation at $k\gtrsim0.1\,h\,{\rm Mpc}^{-1}$ traces the
    displaced baryon-acoustic structure. The horizontal dotted line denotes
    zero response and the vertical dotted line marks the sign change near
    $k_{\rm cross}=0.091\,h\,{\rm Mpc}^{-1}$.}
    \label{fig:matter-power-reshaping}
\end{figure}

The interaction also produces a scale-dependent modification of the linear
matter power spectrum. Figure~\ref{fig:matter-power-reshaping} displays the
fractional response at $z=0$ for the same illustrative fiducial model. The
response reaches a suppression minimum near
$k\simeq0.018\,h\,{\rm Mpc}^{-1}$ and changes sign near
$k\simeq0.0913\,h\,{\rm Mpc}^{-1}$. Representative fractional changes are
approximately
$-9.50\%$ at $k=0.02\,h\,{\rm Mpc}^{-1}$,
$+2.03\%$ at $0.10\,h\,{\rm Mpc}^{-1}$,
$+12.90\%$ at $0.30\,h\,{\rm Mpc}^{-1}$ and
$+20.11\%$ at $1.0\,h\,{\rm Mpc}^{-1}$.

At $k\gtrsim0.1\,h\,{\rm Mpc}^{-1}$, the broad increase in the
linear-theory response is modulated by oscillations associated with the
displacement of baryon-acoustic features in the transfer functions. The close
agreement between the total matter and CDM-plus-baryon responses indicates
that this scale dependence is already present in the clustered matter sector
rather than arising from the distinction between the two matter definitions.
The large-$k$ values shown here are linear theory outputs and should not be
interpreted as predictions in the nonlinear regime.

The examples in this section are intended to expose the physical mechanisms
through which the scalar--photon interaction affects the thermal history,
acoustic structure, lensing and matter clustering. In particular, the
fiducial point in equation~\eqref{eq:fiducial-illustrative-point} was selected for illustrative visibility
rather than statistical preference. The observational analysis is presented
separately in Section~4. We first determine the conditional profile
likelihood in the small-$\epsilon_i$ region and compare the interacting
solutions with a separately evaluated interaction-off $\Lambda$CDM
reference. A complete assessment of parameter degeneracies would require jointly varying the standard cosmological parameters, which is beyond the scope of the present conditional analysis.


\section{Planck likelihood analysis}
\label{sec:likelihood-analysis}

We use the Planck 2018 high-$\ell$ TT, TE and EE lite likelihood,
the low-$\ell$ temperature and polarization likelihoods and the
CMB-lensing likelihood~\cite{planck,planck2018likelihood,planck2018lensing}.
The purpose of the analysis is to characterize the conditional CMB
likelihood response to the dynamical photon evolution described in
Sections~2 and~3, examine its dependence on the interaction parameters,
and test whether the interaction improves the fit relative to the
corresponding interaction-off $\Lambda$CDM cosmology at the adopted fixed
baseline.

Our primary quantitative result is obtained from a conditional
profile-likelihood analysis in the $(\epsilon_i,\sigma)$ plane. In this
calculation the six standard cosmological parameters are fixed at their
adopted baseline values and the Planck calibration parameter
$A_{\rm Planck}$ is profiled. Because the six standard cosmological parameters remain fixed, this calculation cannot establish marginalized parameter constraints or a fully optimized comparison between cosmological models.

Both analyses are performed with MontePython \cite{audren,brinckmann}
interfaced with the modified CLASS implementation described in Section~3.
The results below quantify conditional likelihood responses at the stated fixed baseline. In particular, the conventional $\Delta\chi^2=3.84$ crossing is not interpreted as a general 95\% confidence bound; the fixed parameters may be degenerate with the interaction sector.

For every interacting parameter set, the background calibration is performed
before the recombination history, perturbations, lensed CMB spectra and
lensing-potential spectrum are calculated and passed to the Planck
likelihood.

\subsection{Data sets and likelihood}
\label{subsec:datasets-baseline}

We use the Planck 2018 high-multipole TT, TE and EE lite likelihood,
the low-multipole temperature likelihood, the low-multipole E-mode
polarization likelihood and the CMB-lensing likelihood
\cite{planck2018likelihood,planck2018lensing}. In terms of the likelihood implementation,
the adopted combination consists of high-$\ell$ TTTEEE-lite, low-$\ell$ TT,
low-$\ell$ EE and Planck lensing. The total log-likelihood is written
schematically as
\begin{equation}
\ln{\cal L}_{\rm tot}
=
\ln{\cal L}_{\rm high\text{-}\ell}
+
\ln{\cal L}_{\rm low\text{-}\ell}^{TT}
+
\ln{\cal L}_{\rm low\text{-}\ell}^{EE}
+
\ln{\cal L}_{\rm lens},
\label{eq:total-planck-likelihood}
\end{equation}
and the corresponding effective chi-square is defined by
\begin{equation}
\chi_{\rm eff}^2
\equiv
-2\ln{\cal L}_{\rm tot}.
\label{eq:chi2eff-definition}
\end{equation}

The adopted baseline values of the six standard cosmological parameters are
\begin{align}
\omega_b &\equiv \Omega_b h^2 = 0.02237,
&
\omega_{\rm cdm} &\equiv \Omega_{\rm cdm}h^2 = 0.1200,
\nonumber\\
h &=0.6736,
&
A_s &=2.10\times10^{-9},
\nonumber\\
n_s &=0.9649,
&
\tau_{\rm reio} &=0.0544.
\label{eq:fixed-baseline-parameters}
\end{align}

For the conditional profile analysis, all six quantities in
equation~(\ref{eq:fixed-baseline-parameters}) are held fixed, while
$(\epsilon_i,\sigma)$ are varied and $A_{\rm Planck}$ is profiled.
The treatment of $A_{\rm Planck}$ is kept identical to that used in the
adopted Planck likelihood implementation.

No multidimensional sampling result is used for parameter inference in this work. The fixed-baseline analysis below should therefore be understood as a conditional likelihood sensitivity study.

\subsection{Conditional profile-likelihood methodology}
\label{subsec:profile-methodology}

For a given pair $(\epsilon_i,\sigma)$, the potential slope is fixed by
equation~(\ref{eq:lambda-epsilon-sigma}). The auxiliary background
normalizations are then determined with the shooting procedure described in
Subsection~\ref{subsec:class-implementation-validation}. Only after this
calibration is accepted are the recombination history, CMB spectra and Planck
likelihood evaluated.

The profile likelihood at fixed $(\epsilon_i,\sigma)$ is defined by
maximizing over $A_{\rm Planck}$,
\begin{equation}
{\cal L}_{p}(\epsilon_i,\sigma)
\equiv
\max_{A_{\rm Planck}}
{\cal L}_{\rm tot}
(\epsilon_i,\sigma,A_{\rm Planck};
\boldsymbol{\theta}_{\rm base}),
\label{eq:2d-profile-definition}
\end{equation}
where $\boldsymbol{\theta}_{\rm base}$ denotes the fixed standard
cosmological parameter set specified in
equation~(\ref{eq:fixed-baseline-parameters}). Equivalently, we minimize
$-\ln{\cal L}_{\rm tot}$ with respect to $A_{\rm Planck}$.

The relative profile statistic within the interacting branch is defined as
\begin{equation}
\Delta\chi^2_{\rm prof}(\epsilon_i,\sigma)
\equiv
-2\ln
\left[
\frac{{\cal L}_{p}(\epsilon_i,\sigma)}
     {{\cal L}_{p,\max}^{\rm int}}
\right],
\label{eq:profile-dchi2}
\end{equation}
where ${\cal L}_{p,\max}^{\rm int}$ is the largest profiled likelihood among
the sampled interacting cosmologies. Thus
$\Delta\chi^2_{\rm prof}=0$ at the best sampled interacting point.

Rather than constructing a uniform dense grid, we use an adaptive sequence
of exact evaluations. An initial sparse grid identifies the direction of
increasing likelihood after which additional points refine the
low-$\epsilon_i$ region, the $\Delta\chi^2_{\rm prof}$ crossings and the
dependence on $\sigma$. The final set contains 37 distinct exactly evaluated
interacting cosmologies. The sampled domain spans
\begin{equation}
3\times10^{-5}
\leq
\epsilon_i
\leq
5\times10^{-4},
\label{eq:epsilon-profile-range}
\end{equation}
with $\sigma$ initially explored over
$0.02\leq\sigma\leq0.06$ and subsequently extended to
$\sigma=0.08$ and $0.10$ as a robustness test.

At each cosmology the local dependence on $A_{\rm Planck}$ is sampled around
its minimum, a local quadratic approximation is used to locate the profiled
value and the likelihood is then evaluated explicitly at that value. The
background calibration uses a hard acceptance criterion of $10^{-6}$ for the
largest shooting residual while the final exact profile points were refined
to a target below $5\times10^{-8}$. The largest background residual among
the retained exact cosmologies is
\begin{equation}
r_{\rm max}=4.14\times10^{-8},
\label{eq:profile-max-residual}
\end{equation}
so that numerical calibration errors are well below the likelihood
variations discussed below.

From the two-dimensional profile we also construct the sampled
one-dimensional profiles
\begin{equation}
{\cal L}_{p}(\epsilon_i)
=
\max_{\sigma}{\cal L}_{p}(\epsilon_i,\sigma),
\label{eq:epsilon-profile-definition}
\end{equation}
and
\begin{equation}
{\cal L}_{p}(\sigma)
=
\max_{\epsilon_i}{\cal L}_{p}(\epsilon_i,\sigma).
\label{eq:sigma-profile-definition}
\end{equation}
The maximizations are taken over the available exact sampled values.
These quantities are profile likelihoods, not marginalized probability
distributions, and no Bayesian prior-volume interpretation is attached
to them.

\subsection{Conditional profile-likelihood results}
\label{subsec:profile-results}

\begin{figure}[!htbp]
    \centering
    \includegraphics[width=0.82\textwidth]
    {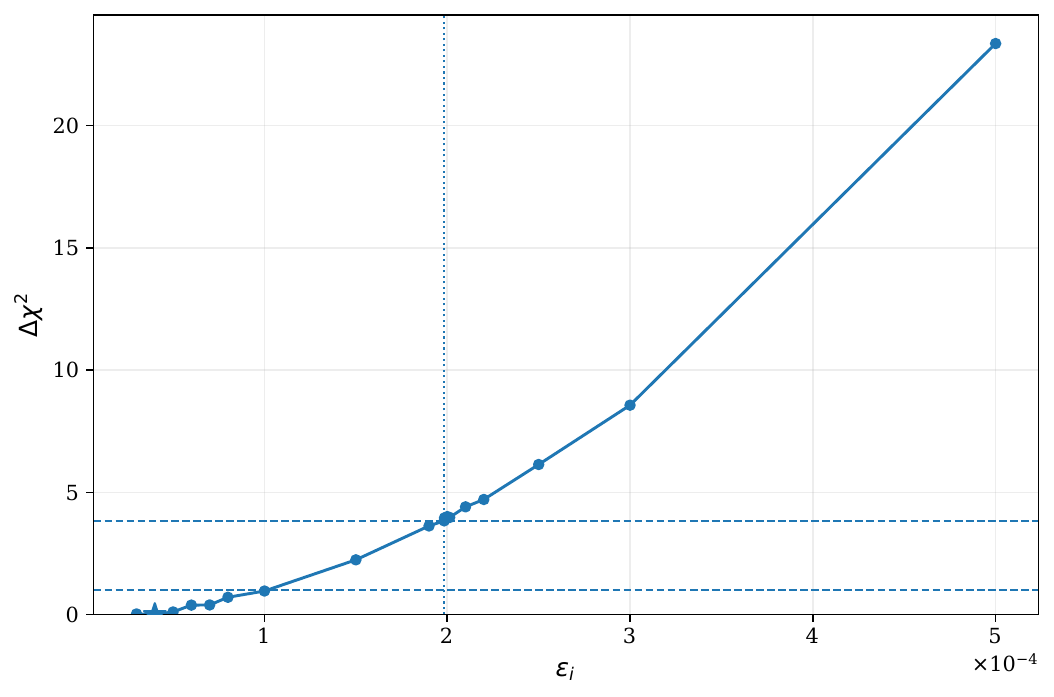}
    \caption{Conditional profile likelihood for the initial radiation era
    scaling parameter $\epsilon_i$. At each sampled $\epsilon_i$, the
    likelihood is maximized over the available values of $\sigma$ while
    $A_{\rm Planck}$ is profiled and the six standard cosmological parameters
    are held fixed at the values in
    equation~(\ref{eq:fixed-baseline-parameters}). The minimum of the sampled
    interacting branch occurs near $\epsilon_i=4\times10^{-5}$. The horizontal
    lines at $\Delta\chi^2_{\rm prof}=1$ and $3.84$ are shown as conventional
    reference levels. The $\Delta\chi^2_{\rm prof}=3.84$ crossing occurs at
    $\epsilon_i\simeq1.98\times10^{-4}$. Because the interaction-off
    $\Lambda$CDM limit is nonregular, these levels should not be interpreted
    automatically as standard frequentist confidence limits.}
    \label{fig:epsilon-profile-likelihood}
\end{figure}

The one-dimensional conditional profile in $\epsilon_i$ is shown in
Figure~\ref{fig:epsilon-profile-likelihood}. The best sampled interacting
cosmology occurs at
\begin{equation}
(\epsilon_i,\sigma)_{\rm int,best}
=
(4.0\times10^{-5},\,0.10),
\label{eq:best-interacting-point}
\end{equation}
with
\begin{equation}
A_{\rm Planck}=1.0004513,
\qquad
-\ln{\cal L}_{\rm int,best}
=
506.480191.
\label{eq:best-interacting-likelihood}
\end{equation}

The conditional minimum is extremely shallow. In particular, the additional
low-$\epsilon_i$ test at
\begin{equation}
(\epsilon_i,\sigma)
=
(3.0\times10^{-5},\,0.10)
\label{eq:low-epsilon-test}
\end{equation}
lies only
\begin{equation}
\Delta\chi^2_{\rm prof}=0.0264
\label{eq:q03-dchi2}
\end{equation}
above the best sampled interacting point. The numerical minimum should
therefore be interpreted as a shallow sampled turnover rather than as
evidence for a preferred non-zero value of $\epsilon_i$.

For orientation, the sampled $\Delta\chi^2_{\rm prof}=1$ level is crossed at
approximately
\begin{equation}
\epsilon_i\simeq1.01\times10^{-4},
\label{eq:dchi2-one-crossing}
\end{equation}
while the more accurately refined conventional
$\Delta\chi^2_{\rm prof}=3.84$ crossing occurs at
\begin{equation}
\epsilon_i
\simeq
1.9844\times10^{-4}.
\label{eq:dchi2-384-crossing}
\end{equation}
The latter value is bracketed by exact evaluations on the
$\sigma=0.10$ slice and is stable under the extension of the sampled
$\sigma$ range. As discussed below,
equation~(\ref{eq:dchi2-384-crossing}) is quoted as a reference profile
crossing rather than as an automatic 95\% confidence bound.

\begin{figure}[!htbp]
    \centering
    \includegraphics[width=0.82\textwidth]
    {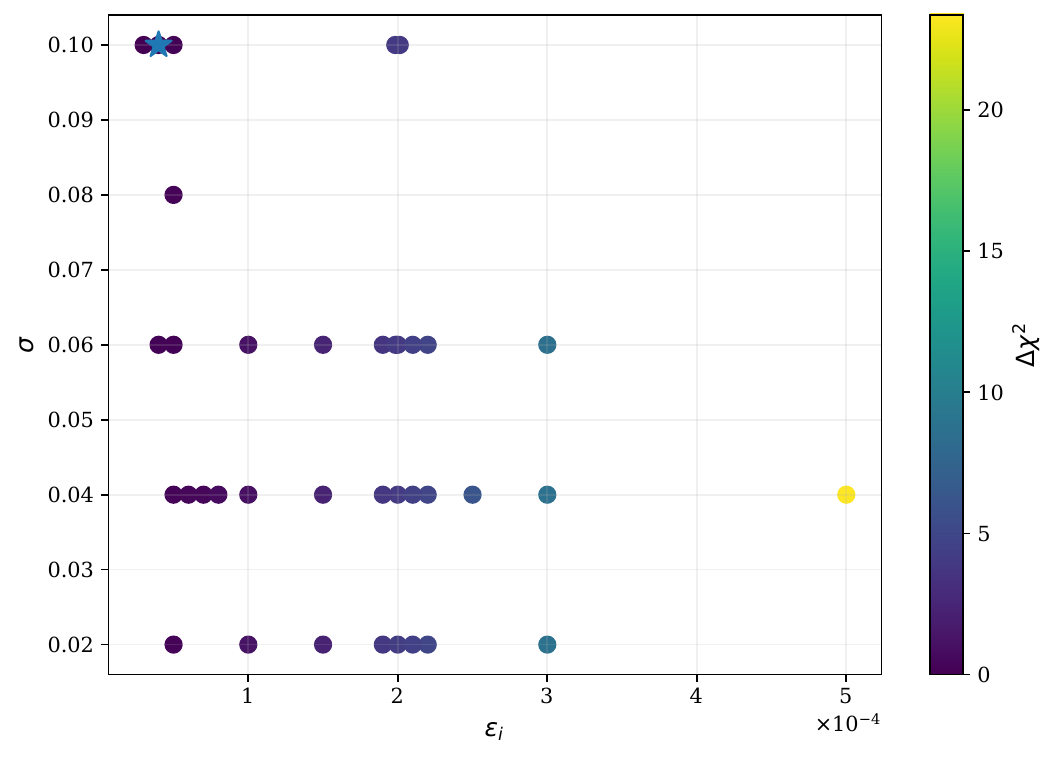}
    \caption{Exact conditional profile-likelihood evaluations in the
    $(\epsilon_i,\sigma)$ plane. The color scale gives
    $\Delta\chi^2_{\rm prof}$ relative to the best sampled interacting point
    after profiling $A_{\rm Planck}$ and holding the six standard
    cosmological parameters fixed. The star marks
    $(\epsilon_i,\sigma)=(4\times10^{-5},0.10)$. The distribution of exact
    points reflects the adaptive sampling strategy: the low-$\epsilon_i$
    region and the conventional profile crossing are sampled more densely,
    while the extension to larger $\sigma$ is used primarily to test
    robustness. The conditional likelihood remains weakly dependent on
    $\sigma$ over the explored range.}
    \label{fig:exact-2d-profile}
\end{figure}

The two-dimensional set of exact evaluations is displayed in
Figure~\ref{fig:exact-2d-profile}. We show the exact points rather than
extrapolating a smooth contour across the more sparsely sampled part of the
extended-$\sigma$ region. The dependence on $\sigma$ is substantially weaker
than that on $\epsilon_i$. Extending the sampled range from
$0.02\leq\sigma\leq0.06$ to $\sigma=0.08$ and $0.10$ does not produce a
meaningful localization of the coupling strength. The occurrence of the
sampled minimum at $\sigma=0.10$ therefore does not constitute evidence for
a preferred value of $\sigma$.

\subsection{Interaction-off $\Lambda$CDM reference and statistical interpretation}
\label{subsec:robustness-lcdm}

The comparison with the interaction-off cosmology requires particular care.
As follows from equation~(\ref{eq:lambda-epsilon-sigma}), the radiation era
scaling parametrization becomes singular as
$\epsilon_i\rightarrow0$. Moreover, once the scalar--photon interaction is
switched off, $\sigma$ no longer labels an observable interaction.
The $\Lambda$CDM reference is therefore evaluated as a separate
interaction-off model rather than by treating $\epsilon_i=0$ as a regular
point within the interacting scaling parametrization.

For this reference calculation, the six standard cosmological parameters are
held at the same baseline values as in the conditional profile analysis and
the same Planck likelihood combination is used. Profiling $A_{\rm Planck}$
gives
\begin{equation}
A_{\rm Planck}^{\Lambda{\rm CDM}}
=
1.000581,
\qquad
-\ln{\cal L}_{\Lambda{\rm CDM}}
\simeq
506.332.
\label{eq:lcdm-profile-result}
\end{equation}
Comparing this value with
equation~(\ref{eq:best-interacting-likelihood}) gives
\begin{equation}
\Delta\chi^2_{\rm int-\Lambda CDM}
\equiv
\chi^2_{\rm eff,int}
-
\chi^2_{\rm eff,\Lambda CDM}
\simeq
+0.296.
\label{eq:int-lcdm-dchi2}
\end{equation}
The positive sign means that the separately evaluated interaction-off
$\Lambda$CDM cosmology has a slightly smaller effective chi-square than the
best sampled interacting cosmology. The difference is very small but there
is therefore no improvement of the Planck fit associated with switching on
the scalar--photon interaction within the conditional analysis.

The nonregular nature of the interaction-off boundary also affects the
interpretation of the one-dimensional profile. In the interaction-off model,
$\sigma$ is not regularly identified and the mapping between
$(\epsilon_i,\sigma)$ and the exponential-potential slope becomes singular as
$\epsilon_i\rightarrow0$. Consequently, the regularity assumptions
underlying the usual Wilks-theorem interpretation of fixed
$\Delta\chi^2$ thresholds are not guaranteed. The crossing
$\Delta\chi^2_{\rm prof}=3.84$ at
$\epsilon_i\simeq1.9844\times10^{-4}$ is therefore quoted only as a
conventional profile-likelihood reference crossing. We do not identify it
automatically with a standard 95\% confidence upper limit, nor do we
translate the small likelihood difference in
equation~(\ref{eq:int-lcdm-dchi2}) into a conventional nested-model detection
significance.

The comparison in equation~(\ref{eq:int-lcdm-dchi2}) applies specifically
to the fixed-baseline analysis. No corresponding multidimensional
optimization of the interacting and interaction-off models has been
performed, so this difference must not be interpreted as a general
model-selection result.

The principal numerical results of the conditional analysis are summarized
in Table~\ref{tab:profile-results}.

\begin{table}[!htbp]
\centering
\begin{tabular}{ll}
\hline
Quantity & Result \\
\hline
Best sampled interacting $\epsilon_i$
& $4.0\times10^{-5}$ \\
Best sampled interacting $\sigma$
& $0.10$ \\
Best sampled interacting $A_{\rm Planck}$
& $1.000451$ \\
Best sampled interacting $-\ln{\cal L}$
& $506.480191$ \\
Low-$\epsilon_i$ test at $3\times10^{-5}$
& $\Delta\chi^2_{\rm prof}=0.0264$ \\
Reference $\Delta\chi^2_{\rm prof}=3.84$ crossing
& $\epsilon_i\simeq1.9844\times10^{-4}$ \\
Interaction-off $\Lambda$CDM $A_{\rm Planck}$
& $1.000581$ \\
Interaction-off $\Lambda$CDM $-\ln{\cal L}$
& $506.332$ \\
$\Delta\chi^2_{\rm int-\Lambda CDM}$
& $+0.296$ \\
\hline
\end{tabular}
\caption{Summary of the conditional Planck profile-likelihood analysis.
The low-$\epsilon_i$ likelihood difference is quoted relative to the best
sampled interacting cosmology. The $\Lambda$CDM likelihood is obtained from
a separate interaction-off calculation rather than by setting
$\epsilon_i=0$ within the interacting scaling parametrization. The stored
$\Lambda$CDM value of $-\ln{\cal L}$ is available to approximately
$10^{-3}$ precision and no additional significance should be attached to
digits beyond that level.}
\label{tab:profile-results}
\end{table}

\section{Discussion and conclusions}
\label{sec:discussion-conclusions}

We have investigated the cosmological consequences of an effective coupling
between a canonical scalar field and the photon fluid. Unlike a prescribed,
constant modification of the photon dilution law, the parameter $\epsilon_i$
selects a radiation-era scaling trajectory used for initialization, while the
subsequent interaction history follows the coupled field equations. This
distinction matters for recombination: the photon-temperature history and the
expansion rate respond to the evolving scalar field, and both affect the
visibility function and acoustic structure.

The CLASS implementation provides a framework for following these responses
through the background and linear perturbations. The illustrative fiducial
solution in Section~\ref{sec:numerical-results} shows changes in the
recombination history, CMB anisotropies, lensing, and the linear matter power
spectrum. The comparatively large fiducial interaction was chosen to reveal
these effects; it is not an observationally preferred parameter choice.
Archived precision-setting tests support the stability of the principal CMB
spectra through $\ell=3000$. A separate tight-coupling derivative check
changes the TT spectrum by at most approximately $2.4\times10^{-6}$ in
relative amplitude over $30\leq\ell\leq2500$ at the tested weakly
interacting point. This is a limited numerical consistency check, not a
validation of the full photon kinetic hierarchy. A separate
matched-initialization-redshift test has not been established from the
available archive, and we do not claim that it has been completed.

In the conditional Planck 2018 comparison, the best sampled interacting
point is $(\epsilon_i,\sigma)=(4\times10^{-5},0.10)$. The nearby point
$(3\times10^{-5},0.10)$ differs by only
$\Delta\chi^2_{\rm prof}=0.0264$, indicating a shallow likelihood response.
The separately evaluated interaction-off $\Lambda$CDM reference is slightly
better than the best sampled interacting solution,
$\Delta\chi^2_{\rm int-\Lambda CDM}\simeq+0.296$. Thus, within the
specified fixed-baseline comparison, the interaction does not improve the
Planck likelihood. These differences should not be interpreted as evidence
for a preferred non-zero coupling or as a global comparison between fully
optimized cosmological models.

The conventional $\Delta\chi^2_{\rm prof}=3.84$ crossing near
$\epsilon_i\simeq1.9844\times10^{-4}$ is retained only as a reference
feature of the conditional profile. The six standard cosmological parameters
were not varied, and the interaction-off limit is nonregular: the potential
slope parametrization becomes singular as $\epsilon_i\rightarrow0$ and
$\sigma$ is not identified when the coupling is absent. Consequently, a
standard Wilks-theorem confidence-level interpretation is not established
\cite{wilks1938}. A defensible cosmological constraint requires matched
multidimensional profiling or converged sampling with the standard parameters
allowed to vary.

There are also physical qualifications that are independent of statistical
precision. The unweighted thermal photon density $\rho_\gamma$ differs from
the gravitational source $e^{-\sigma\phi}\rho_\gamma$. Matching the
present-day thermal density does not, on its own, determine the present-day
coupling normalization. Moreover, the modified temperature law follows from
the photon continuity equation only after assuming that the photon bath
remains Planckian. The effective perfect-fluid action does not specify the
microscopic photon-number-changing or thermalization processes required to
assess spectral distortions, nor does it uniquely fix the higher photon
multipoles and scattering collision terms used in the phenomenological
Boltzmann implementation. A quantitative confrontation with the observed
near-blackbody CMB spectrum therefore remains outstanding; the present work
does not claim a spectral-distortion bound.

Independent observations of the CMB temperature--redshift relation and
cosmic distance duality provide complementary tests of non-standard photon
evolution \cite{ruchika2025,martins2026,wang2024cddr}, but they are not
translated here into direct limits on $(\epsilon_i,\sigma)$. Our results
establish the signatures of the effective model within its stated
thermodynamic assumptions and show that a weak interaction can remain close
to the standard CMB prediction at the adopted cosmological baseline.
The next substantive advances are to validate the photon normalization and
initialization independence, provide a kinetic treatment of photon
thermalization, and perform a joint cosmological parameter analysis. These
requirements limit the strength of any observational interpretation of the
current calculations.

\paragraph{Data and code availability.}
The observational data used in this work are publicly available through the
corresponding survey and likelihood releases. The modified CLASS and
MontePython configuration files, plotting scripts and numerical chains can be
made available by the author upon reasonable request.

\paragraph{Acknowledgments.}
This work is supported by Shahid Rajaee Teacher Training University under
grant No.~23905/380001.

\appendix

\section{Perturbation sources}
\label{app:perturbation-sources}

For completeness, we collect here the scalar field and photon contributions
to the linearized Einstein equations. For the canonical scalar field, the
density, pressure and momentum perturbations are
\begin{subequations}
\label{eq:scalar-fluid-perturbations}
\begin{align}
\delta\rho_\phi
&=
\frac{\bar\phi'\delta\phi'}{a^2}
+
V_{,\phi}\delta\phi,
\\
\delta p_\phi
&=
\frac{\bar\phi'\delta\phi'}{a^2}
-
V_{,\phi}\delta\phi,
\\
(\bar\rho_\phi+\bar p_\phi)\theta_\phi
&=
\frac{k^2\bar\phi'\delta\phi}{a^2}.
\end{align}
\end{subequations}

The photon stress-energy tensor enters the Einstein equations multiplied by
the explicit factor $e^{-\sigma\phi}$. Perturbing this weighted photon
contribution gives
\begin{subequations}
\label{eq:weighted-photon-sources}
\begin{align}
\delta\!\left(e^{-\sigma\phi}\rho_\gamma\right)
&=
e^{-\sigma\bar\phi}\bar\rho_\gamma
\left(\delta_\gamma-\sigma\delta\phi\right),
\\
\delta\!\left(e^{-\sigma\phi}p_\gamma\right)
&=
\frac{1}{3}
e^{-\sigma\bar\phi}\bar\rho_\gamma
\left(\delta_\gamma-\sigma\delta\phi\right),
\\
e^{-\sigma\bar\phi}
(\bar\rho_\gamma+\bar p_\gamma)\theta_\gamma
&=
\frac{4}{3}
e^{-\sigma\bar\phi}\bar\rho_\gamma\theta_\gamma,
\\
e^{-\sigma\bar\phi}
(\bar\rho_\gamma+\bar p_\gamma)\Pi_\gamma
&=
\frac{4}{3}
e^{-\sigma\bar\phi}\bar\rho_\gamma\Pi_\gamma.
\end{align}
\end{subequations}

The combination $\delta_\gamma-\sigma\delta\phi$ in the density and pressure
sources arises from perturbing the explicit factor $e^{-\sigma\phi}$.
No analogous linear correction from the perturbed prefactor appears in the
momentum or anisotropic-stress terms because the corresponding background
photon momentum and shear vanish.

For the isotropic energy transfer closure adopted in the numerical
implementation, there is no direct interaction force in the photon Euler
equation. The modified photon monopole source in
equation~(\ref{eq:photon-continuity-coupled}) subsequently propagates through
the usual free streaming and Thomson-scattering hierarchy.

\section{Derivation of the adiabatic initial conditions}
\label{app:adiabatic-initial-conditions}

We summarize the leading super-Hubble adiabatic relations used to motivate
the initial conditions of Section~2.4. Their validity as a complete regular
growing-mode solution beyond leading order is not established by the
time-shift argument alone. On the radiation era scaling branch characterized
by $\epsilon_i$,
\begin{equation}
\bar\phi'
=
\frac{3\epsilon_i}{4\sigma}{\cal H},
\qquad
\bar\rho_\gamma'
=
-(4-\epsilon_i){\cal H}\bar\rho_\gamma.
\label{eq:initial-background-scaling}
\end{equation}

The photon contribution appearing in the Einstein equations satisfies
\begin{equation}
\left(
e^{-\sigma\bar\phi}\bar\rho_\gamma
\right)'
=
-\left(4-\frac{\epsilon_i}{4}\right)
{\cal H}
e^{-\sigma\bar\phi}\bar\rho_\gamma.
\label{eq:initial-weighted-photon-scaling}
\end{equation}
This follows directly from the radiation-era scaling solution together with
the explicit $e^{-\sigma\bar\phi}$ weighting.

For the regular growing adiabatic mode, all components correspond to the same
local time shift,
\begin{equation}
\frac{\delta\rho_i}{\bar\rho_i'}
=
\frac{\delta\rho_j}{\bar\rho_j'}.
\label{eq:general-adiabatic-condition}
\end{equation}
The scalar trajectory obeys the same time-shift condition. Adiabaticity
between the scalar field and the photon fluid therefore gives
\begin{equation}
\frac{\delta\phi}{\bar\phi'}
=
\frac{\delta\rho_\gamma}{\bar\rho_\gamma'}
=
-\frac{\delta_\gamma}
       {(4-\epsilon_i){\cal H}},
\label{eq:scalar-photon-timeshift}
\end{equation}
and hence
\begin{equation}
\delta\phi
=
-\frac{3\epsilon_i}
       {4\sigma(4-\epsilon_i)}
\delta_\gamma.
\label{eq:scalar-photon-adiabatic-relation}
\end{equation}

Using equation~(\ref{eq:scalar-photon-adiabatic-relation}), the perturbation
of the weighted photon contribution satisfies
\begin{equation}
\frac{\delta_\gamma-\sigma\delta\phi}
     {4-\epsilon_i/4}
=
\frac{\delta_\gamma}{4-\epsilon_i}
=
\frac{\delta_\nu}{4}
=
\frac{\delta_b}{3}
=
\frac{\delta_c}{3}.
\label{eq:weighted-photon-adiabatic-relation}
\end{equation}
Thus the photon density contrast evolved by the Boltzmann hierarchy and the
weighted photon source entering the Einstein equations correspond to the same
adiabatic time-shift mode. No additional photon-fluid variable is required.

For the leading synchronous-gauge growing mode
\begin{equation}
h=C(k\tau)^2,
\label{eq:appendix-growing-mode}
\end{equation}
the photon-velocity contribution to
equation~(\ref{eq:photon-continuity-coupled}) is of higher order in $k\tau$.
Combining that equation with
equation~(\ref{eq:scalar-photon-adiabatic-relation}) gives the initial density
and scalar amplitudes quoted in
equations~(\ref{eq:initial-density-conditions}) and
(\ref{eq:initial-scalar-conditions}).

In particular, the weighted photon density source initially obeys
\begin{equation}
\delta_{\gamma i}
-
\sigma\delta\phi_i
=
-\frac{4-\epsilon_i/4}{6}h_i.
\label{eq:initial-weighted-photon-condition}
\end{equation}
The derivative relation in
equation~(\ref{eq:initial-scalar-conditions}) is the leading super-Hubble
result on the radiation-era scaling branch. Finite-starting-time corrections
are checked as part of the numerical validation described in Section~3.

\section{Supplementary CMB responses}
\label{app:supplementary-cmb-responses}

\begin{figure}[!htbp]
    \centering
    \includegraphics[width=0.82\textwidth]
    {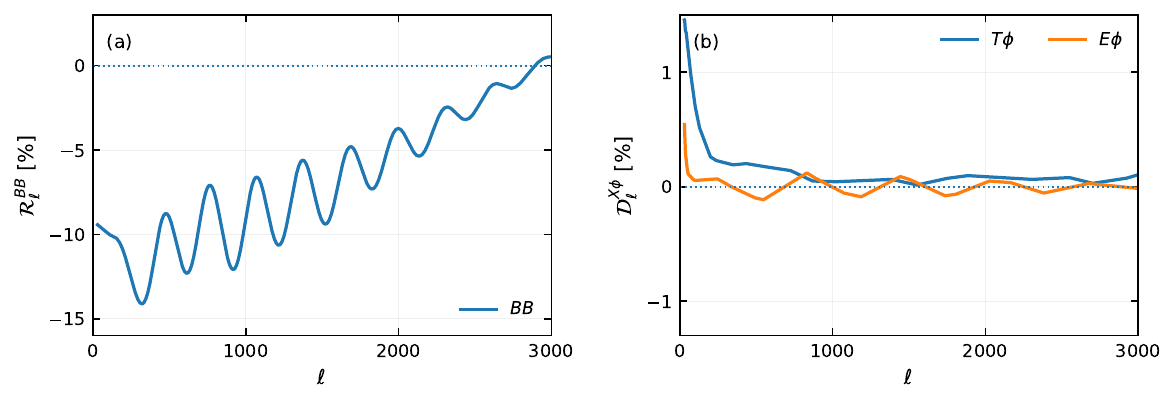}
    \caption{Supplementary lensed-CMB responses of the illustrative fiducial
    scalar--photon model relative to the matched interaction-off
    $\Lambda$CDM reference. Panel (a) shows the BB auto-spectrum response
    ${\cal R}_{\ell}^{BB}$ while panel (b) shows the normalized
    $T\phi$ and $E\phi$ cross-spectrum differences. These results complement
    the principal TT, EE, TE and lensing responses shown in the main text.}
    \label{fig:supplementary-cmb-residuals}
\end{figure}

The main text displays the TT, EE, TE and lensing-potential responses that
most directly expose the displacement of the acoustic pattern and the change
in the integrated lensing potential. For completeness,
Figure~\ref{fig:supplementary-cmb-residuals} shows the complementary
BB, $T\phi$ and $E\phi$ responses for the same illustrative fiducial model.

The BB response arises primarily through the lensing remapping of E-mode
polarization, whereas the temperature--lensing and polarization--lensing
cross-spectrum responses remain comparatively small for this parameter
choice. As in Section~3, these curves are fixed parameter physical response
tests and are not statistical constraints.

\end{document}